\documentclass{aa}  

\usepackage{graphicx}
\usepackage{txfonts}
\usepackage[colorlinks=true,
    linkcolor=blue,
    citecolor=blue,
    filecolor=magenta,      
    urlcolor=cyan]{hyperref}
\usepackage{booktabs}
\usepackage{natbib}
\usepackage{caption}
\usepackage{subcaption}
\usepackage{txfonts}
\usepackage[normalem]{ulem}
\usepackage[dvipsnames]{xcolor}

\begin{document}

   \title{Black hole merger rates for LISA and LGWA from semi-analytical modelling of light seeds}
   \titlerunning{Black hole merger rates for LISA and LGWA}
   \authorrunning{J. Singh et al.}

   \author{Jasbir Singh
          \inst{1}\fnmsep\thanks{jasbir.singh@inaf.it},
          Paola Severgnini\inst{1},
          Vieri Cammelli\inst{2},
          Alessandra De Rosa\inst{3},
          Cristian Vignali\inst{4,5},
          Fabio Rigamonti\inst{1,6,7},
          Rosa Valiante\inst{8},
          Pierluigi Monaco\inst{9,10,11,12},
          Jonathan C. Tan\inst{13,14},
          Lorenzo Battistini\inst{3, 15},
          Roberto Della Ceca\inst{1},
          Jan Harms\inst{16,17},
          \and
          Manali Parvatikar\inst{3,18}
          }
   
   \institute{
            INAF - Osservatorio Astronomico di Brera, via Brera 20, I-20121 Milano, Italy
        \and
            Department of Physics, Informatics \& Mathematics, University of Modena \& Reggio Emilia, via G. Campi 213/A, 41125, Modena, Italy
        \and
            INAF - Istituto di Astrofisica e Planetologia Spaziali (IAPS), via Fosso del Cavaliere, Roma, I-133, Italy
        \and
            Dipartimento di Fisica e Astronomia `Augusto Righi’, Università degli Studi di Bologna, Via Gobetti 93/2, I-40129 Bologna, Italy
        \and
            INAF – Osservatorio di Astrofisica e Scienza dello Spazio di Bologna, Via Gobetti 93/3, I-40129 Bologna, Italy
        \and
            INFN, Sezione di Milano-Bicocca, Piazza della Scienza 3, I-20126 Milano, Italy
        \and
            Como Lake centre for AstroPhysics (CLAP), DiSAT, Università dell’Insubria, via Valleggio 11, 22100 Como, Italy
        \and
            Istituto Nazionale di Astrofisica, Osservatorio Astronomico di Roma, Via Frascati 33, 00077 Monte Porzio Catone, Italy
        \and
            Dipartimento di Fisica, Sezione di Astronomia, Università degli Studi di Trieste, via G.B. Tiepolo 11, I-34131, Trieste, Italy
        \and
            INAF - Osservatorio Astronomico di Trieste, via G.B. Tiepolo 11, I-34131, Trieste, Italy
        \and
            INFN, Sezione di Trieste, Via Valerio 2, 34127 Trieste TS, Italy
        \and
            IFPU, Institute for Fundamental Physics of the Universe, via Beirut 2, 34151 Trieste, Italy
        \and
            Department of Space, Earth \& Environment, Chalmers University of Technology, SE-412 96 Gothenburg, Sweden
        \and
            Department of Astronomy, University of Virginia, Charlottesville, VA 22904-4235, USA
        \and
            Dipartimento di Matematica e Fisica, Università Roma Tre, via della Vasca Navale 84, I-00146 Roma, Italy
        \and
            Gran Sasso Science Institute (GSSI), I-67100 L’Aquila, Italy
        \and
            INFN, Laboratori Nazionali del Gran Sasso, I-67100 Assergi, Italy
        \and
            Dipartimento di Fisica, Università di Roma Tor Vergata, Via della Ricerca Scientifica, I-00133, Roma, Italy
             }

   % \date{Received September 15, 1996; accepted March 16, 1997}

% \abstract{}{}{}{}{} 
% 5 {} token are mandatory

  \abstract
  % context heading (optional)
  % {} leave it empty if necessary
   {With the upcoming space- and Moon-based gravitational-wave detectors, LISA and LGWA respectively, a new era of GW astronomy will begin with the possibility of detections of the mergers of intermediate-mass black holes (IMBHs) and supermassive black holes (SMBHs).}
  % aims heading (mandatory)
   {We generate populations of synthetic black hole (BH) binaries with masses ranging from the intermediate ($10^3-10^5 M_\odot$) to the supermassive regime ($>10^5 M_\odot$), formed from the dynamical processes of merging halos and their residing galaxies, assuming that each galaxy is initially seeded with a single black hole at its centre. The aim is to estimate the rate of these BH mergers which could be detected by LISA and LGWA.}
  % methods heading (mandatory)
   {Using \textsc{pinocchio} cosmological simulation and a semi-analytical model based on GAEA, we construct a population of merging BHs by implementing a "light" seeding scheme and calculating the merging timescales using the Chandrasekhar prescription. We provide upper and lower limits of dynamical friction timescale by varying the mass of the infalling object to create "pessimistic" and "optimistic" merger rates respectively.}
  % results heading (mandatory)
   {We find that for our synthetic population of MBHs, both LGWA and LISA are able to detect more than $15$ binary IMBH mergers per year in the optimistic case, while in the pessimistic case less than $\sim5$ detections would be possible considering the entire lifetime of the detectors. For SMBHs, the rates are slightly lower in both cases. Most mergers below $z\approx4$ are detected in the optimistic case, although mergers beyond $z=8$ are also detectable at a lower rate.}
  % conclusions heading (optional), leave it empty if necessary 
   {We find that LGWA is better suited for high-SNR IMBH detections at higher redshift, while LISA is more sensitive to massive SMBHs. Joint observations will probe the full BH mass spectrum and constrain BH formation and seeding models.}


   \maketitle
%
%-------------------------------------------------------------------

\section{Introduction}
\label{sec:intro}

Since the first direct detection of a binary black hole (BH) merger through Gravitational Waves (GW) by the Laser Interferometer Gravitational-wave Observatory (LIGO) Collaboration in 2015  \citep{LIGO15, Abbott16}, the field of GW astronomy has grown rapidly. Multiple binary BH mergers have been detected, with masses reaching as high as $\sim 225 M_\odot$ from the latest observational run O4a \citep{LIGO25}. These observations provide a proof of the existence of stellar BHs, defined with masses below $\mathcal{O}(100) M_\odot$. On the other side of the mass spectrum, BHs with masses larger than $\sim10^5 M_\odot$, i.e. supermassive black holes (SMBHs), are well established through electromagnetic observations. These objects reside in the centres of massive galaxies and, when active, power Active Galactic Nucleus \citep[AGN, e.g.][]{Ferrarese05,Volonteri10,Inayoshi20,Lusso23,Harikane23,Lupi24}. They have also been observed directly by the Event Horizon Telescope Collaboration (\citeyear{EHT19,EHT22}), who imaged the SMBHs at the centre of Messier 87 galaxy and our Milky Way galaxy. 
In addition to electromagnetic observations, the mergers of SMBHs are expected to produce low-frequency GWs detectable at the nano-hertz frequencies by the Pulsar Timing Arrays \citep[PTA,][]{Agazie23,Agazie23b,EPTA23,Reardon23,Xu23} experiments. 
On the other hand, the mergers at relatively higher frequencies (milli- and deci-hertz ranges) of both SMBHs and intermediate-mass black holes (IMBHs) will be potentially detected by instruments such as the upcoming space-based Laser Interferometer Space Antenna \citep[LISA,][]{Amaro-Seoane13,LISA23}, and the Moon-based Lunar Gravitational-wave Antenna \citep[LGWA,][]{Harms21,Ajith25}. In this work, we will focus on these two detectors, although we note that additional detectors have been proposed, namely, the Laser Interferometer Lunar Antenna \citep[LILA,][]{Jani25}, TianQin \citep{Luo16, Luo25}, the Deci-hertz Interferometer Gravitational Wave Observatory \citep[DECIGO,][]{Kawamura21}, among others.

IMBHs, with masses in the range $\sim10^3$ to $\sim10^5$ $M_\odot$, occupy  the middle of the BH mass spectrum and remain one of the most elusive BH population, with limited observational evidence \citep[see][for review]{Greene20}. Confirming their existence remains an open question in astrophysics and one promising way is the detection of GWs generated by their mergers.

Within the framework of hierarchical galaxy evolution, galaxies merge, and those hosting massive black holes (MBHs; bracketing both IMBHs and SMBHs) at their centre, would be brought into close proximity, potentially leading to a merger. The masses of BHs at the centres of merging galaxies can span from intermediate to supermassive scales by the time they merge, depending on factors like initial seed mass, accretion history, and prior mergers with compact objects such as neutron stars or other BHs.

The initial seed mass itself can arise from different formation channels, which are broadly categorized into “light” seed and “heavy” seed scenarios. The former is based on the collapse of Pop III stars, the first generation of stars formed in environments with little or no metal content. Conventional models of Pop III star formation predict stars forming with Initial Mass function (IMF) peaking at $\sim100 M_\odot$, but with a tail extending to $\sim10^3 M_\odot$ \citep[e.g.][]{Madau01,Abel02,Tan04,Greif2006,McKee08,Tan10}. Once these stars collapse, they form IMBHs, which might grow in mass due to mergers and accretion. IMBHs are shown to also form in dense stellar clusters through runaway stellar mergers, repeated BH mergers, and runaway Tidal Disruption Events \citep[e.g.][]{Ebisuzaki03,Zwart04,Giersz15,ArcaSedd18,Antonini19,Rizzuto23}. However, due to the challenges in resolving the formation and evolution of individual stars inside clusters required to estimate the formation of IMBHs inside galaxies, the predictions for the cosmological population of these IMBH populations are highly uncertain \citep[see, e.g.][]{Boekholt18,Chon20,Tagawa20}.

The heavy seed formation mechanisms, on the other hand, produce BH seeds of the order of $\sim10^4$ to $\sim10^6 M_\odot$. Although the exact mechanism of their formation is not known, the most popular theory is the "direct collapse" scenario. This process involves the collapse of a massive primordial gas cloud in an atomically-cooled halo of $\sim10^8 M_\odot$, leading to the formation of a single supermassive star of $10^4-10^6 M_\odot$, which then collapses to form an SMBH \citep[e.g.][]{Bromm03,Shang10,Maio19,Bhowmick22a}. However, the alignment of required conditions such as reduced fragmentation, shorter growth timescales, suppression of molecular hydrogen, and more, makes this mechanism relatively rare and unable to explain the entire population of SMBHs \citep{Latif13,Habouzit16,Chon16,Wise19,Haemmerle20}. Other heavy seed scenarios, such as the Pop III.1 model \citep{Banik19,Singh23,Cammelli25}, which is capable of explaining the entire population of SMBHs, the model presented in \citet{Feng21}, which accounts for the origin of high-redshift SMBHs, and the Primordial Black hole formation scenarios \citep[e.g.][]{Hawking71,Carr05,Dayal24}, among others, require exotic physics. In contrast to the heavy seed scenario, through which only SMBHs can be produced, the light seed scenario allows for the formation of smaller seeds, which can grow into IMBHs and SMBHs. Therefore, galaxies seeded with BHs from the collapse of Pop III stars at their centre would later merge, producing IMBH binaries, SMBH binaries, and also IMBH-SMBH binaries.

Due to the paucity of observational constraints on the IMBH population, synthetic IMBH populations generated by large-scale cosmological simulations or semi-analytical models (SAMs) largely depend on the assumptions on the formation and growth of the BHs. Simulations or SAMs which seed dark matter (DM) halos or galaxies with heavy seeds naturally have a lower limit on the mass of MBH which could be anywhere ranging from $\sim10^4 M_\odot$ to $10^6 M_\odot$. For instance, the Halo Mass Threshold (HMT) scheme - based on the methods developed by \citet{Sijacki07} and \citet{DiMatteo08}, and implemented in the Illustris project \citep{Vogelsberger14} - seeds a BH of mass $1.4\times10^5 M_\odot$ in every halo that exceeds a mass threshold of $m_\mathrm{th}=7.1\times10^{10} M_\odot$. The Horizon-AGN simulation \citep{Volonteri16} adopts a more selective approach, taking into account additional physical properties of the host galaxy when determining BH seeding. In this model, a BH seed with a mass of $10^5M_\odot$ is introduced only if the galaxy exceeds specific thresholds in gas density, stellar density, and stellar velocity dispersion. By contrast, simulations and SAMs using "light seeds" have BHs as low as tens of solar masses, which effectively implement a Pop III scenario, producing both IMBH and SMBH at lower redshifts. For example, the SAM Cosmic Archaeology Tool \citep[\texttt{CAT},][]{Trinca22,Valiante16} and \texttt{L-Galaxies} \citep{Izquierdo-Villalba20,Izquierdo-Villalba22,Spinoso23} both have BH seed masses ranging from $\sim10\:M_\odot$ to $10^5 M_\odot$. This results in a BH population at low redshift that spans an extremely broad mass range, largely unconstrained by observations. Furthermore, due to diverse assumptions and physical processes included in these simulations and SAMs to model the mergers of DM halos, the galaxies and finally the BHs, the predicted merger rate of BHs ends up spanning a huge range, going from as low as less than 1 merger per year to more than 100 \citep[for a review of different models see][]{Amaro-Seoane23}.

As mentioned above, LISA and LGWA will be able to detect GWs from mergers involving both IMBHs and SMBHs \citep{Amaro-Seoane23,Colpi24,Ajith25}. LISA, expected to launch around 2035, will detect GWs in the $10^{-4}$ to 1 Hz range, with peak sensitivity near a few millihertz. It can observe the full inspiral, merger, and ringdown of binaries with total masses between $10^5-10^7\:M_\odot$, and the inspiral phase for $10^3-10^4\:M_\odot$ binaries \citep{Colpi24}. LISA is most sensitive to $10^5-10^6\:M_\odot$  binaries, with detection capabilities up to $z\sim20$ \citep{Valiente21,Amaro-Seoane23}. 
LGWA on the other hand, consists of seismic sensors placed on the surface of the Moon in order to detect GWs passing through \citep{Harms21,Ajith25}. It will cover mHz to a few Hz, peaking in the decihertz band. While overlapping with LISA, LGWA offers higher signal-to-noise ratio (SNR) for IMBH binaries. Both detectors can observe binaries in the $10^3-10^6\:M_\odot$ range, enabling joint detections and improved parameter estimation \citep{Ajith25}. Furthermore, any detection of the merger of binaries of IMBH will provide crucial constraints on their population, while also shedding light on the binary formation processes.

In this work, we estimate the detection rates of MBH binary mergers by LISA and LGWA detectors, focusing on the potential to observe IMBHs. We implement the formation channel of IMBHs and SMBHs which involves seeding each galaxy with a single BH at its centre. In Section \ref{sec:simulation}, we present the cosmological simulation and the SAM used to generate the population of BH binaries. The methodology for constructing the population of binaries is presented in Section \ref{sec:methods}. In Section \ref{sec:results}, we show the evolution of merger rates convolved with the sensitivity of the GW detectors across different total binary mass ranges. We discuss the implications of these results for the detectability of IMBHs and the synergies between LISA and LGWA, along with a comparison of our results with the literature. Section \ref{sec:conclusions} summarises our conclusions.

\section{Simulations}
\label{sec:simulation}

\subsection{The dark matter skeleton}

To obtain the merger trees of a representative set of DM halos, we use version 5 of the \textsc{pinocchio} \citep[PIN-pointing Orbit Crossing-Collapsed HIerarchical Objects][]{Monaco02,Munari17} code, described in \cite{Euclid25}. This code is based on Lagrangian Perturbation Theory \citep[e.g.][]{Moutarde91} and works with the same settings as an $N$-body simulation, but at a tiny fraction ($\sim$ a thousandth) of the computational cost. It produces catalogues of DM halos with information on their mass, position, and velocity. The code also generates for each halo a merger history with continuous time sampling. Every time two halos merge, the larger one retains its identity while the smaller one disappears; this means that no information is retained on the internal structure of the halo. This is the main reason for the dramatic speed-up of the code with respect to an $N$-body simulation, where most time is spent in integrating orbits in the highest density regions. However, as shown in \cite{Cammelli25}, information on the substructure can be recovered by modelling the gradual disruption of the smaller halo orbiting inside the main one.

For this work, we utilised the same simulation box used in \cite{Singh23} and \cite{Cammelli25}, which is a box of 59.7 Mpc (40 Mpc/h for $h=0.67$) side length with standard Planck Cosmology (\citeyear{Planck20}), simulated down to redshift $z=0$. The mass of each DM particle is $7.9\times10^6 M_\odot$, and each DM halo is resolved with 10 particles, i.e. the minimum mass of the halo is $7.9\times10^7 M_\odot$. 

\subsection{The semi analytical framework}

To predict galaxy properties based on host DM halos derived form \textsc{pinocchio}, we make use of the semi-analytic approach as presented in \citet{Cammelli25} (hereafter CAM25). In particular, CAM25 introduces a combined model based on DM halo merger trees extracted from \textsc{pinocchio} and uses the SAM GAEA \citep{Hirschmann16, Fontanot20, DeLucia24} to populate each halo with synthetic galaxies by implementing both theoretically and empirically derived prescriptions for star formation, BH accretion and their feedbacks on the halo and galaxy gas components. It is worth noting that in a semi-analytical framework as CAM25, each galaxy is made of idealised bulge and disc components that can host at most one central BH (the presence of wandering BHs is neglected). The CAM25 combined model is calibrated to reproduce galaxy luminosity functions and scaling relations down to redshift $z=0$ for galaxies hosted in DM halo masses as low as $10^8 M_\odot$. SMBH accretion prescriptions are tuned to match AGN luminosity functions up to $z\sim4$, as described in \citet{Fontanot20}. Considering these calibrations at relatively low redshift, and the large uncertainties in both theoretical and observational constraints on early galaxy formation, we note that the current implementation of the CAM25 model may not be optimal for accurately tracking and capturing the first phases of galaxy/BH formation and evolution. In particular, while the model may be missing accurate recipes for early BH accretion, star formation (e.g. Pop III stars), AGN and SNa feedback, we stress that the lack of knowledge and/or observational constraints at earlier redshifts does not allow us to rely on a fiducial high redshift physical treatment. We opt instead for a more conservative choice, keeping the model parametrization constrained to lower redshift scaling relations. In such a way, we preserve the predictive power of the model by using parameters which reproduce observed properties in the local Universe. In the next subsection, we briefly summarise some of the prescription adopted in CAM25 which are most relevant for this work, i.e. BH accretion and galaxy sizes. We refer to CAM25 and the references therein for a full description of the model.

\subsubsection{Black hole accretion}

In CAM25, accretion onto BHs follows two primary modes, detailed in \citet{Fontanot20}. The first channel is the radio-mode, where the accretion rate ($\dot{M}_{\text{R}}$) is linearly proportional to the mass of the BH ($M_{\rm BH}$), virial velocity ($V_{\rm vir}$) to the cube, and hot gas fraction ($f_{\rm hot}$), scaled by a free parameter $k_{\rm radio}$:
\begin{equation} \label{eq:radio_mode}
    \dot{M}_{\text{R}} = k_{\rm radio} \frac{M_{\rm BH}}{10^8 M_{\odot}} \frac{f_{\rm hot}}{0.1} \Bigg(\frac{V_{\rm vir}}{200\:{\rm km/s}}\Bigg)^3.
\end{equation}

The second channel, mostly contributing to the BH mass build-up, is the QSO-mode, which models cold gas accretion onto SMBHs. This mode involves a three-phase process: 1) cold gas losing angular momentum and forming a central reservoir, triggered by disc instabilities or galaxy mergers; 2) gas from this reservoir accreting onto the BH; and 3) resulting AGN-driven outflows expelling gas. The reservoir growth is proportional to the central star formation rate ($\psi_{\rm cs}$) during mergers:
\begin{equation} \label{eq:merg}
    \dot{M}_{\rm rsv}^{\rm cs} = f_{\rm lowJ} \psi_{\rm cs},
\end{equation}
or to the bulge growth rate ($\dot{M}_{\rm bulge}$) during disc instabilities:
\begin{equation}\label{eq:disc}
    \dot{M}_{\rm rsv}^{\rm di} = f_{\rm lowJ}\, \mu\, \dot{M}_{\rm bulge}.
\end{equation}

In both cases the actual accretion is adjusted via two free parameters, the fraction of gas accreted via angular momentum loss $f_{\rm lowJ}$ and its relative fraction sinking into the bulge $\mu$ (values as reported in CAM25).

Once the gas is in the reservoir, the viscosity of the gas leads to an accretion rate onto the BH ($\dot{M}_{\rm BH}$) which is determined by:
\begin{equation} \label{eq:rate_bh}
    \dot{M}_{\rm BH} = f_{\rm BH} \frac{\sigma_{\rm B}^3}{G} \Bigg(\frac{M_{\rm rsv}}{M_{\rm BH}}\Bigg)^{3/2} \Bigg(1 + \frac{M_{\rm BH}}{M_{\rm rsv}}\Bigg)^{1/2},
\end{equation}
where the bulge velocity dispersion is represented by $\sigma_{\rm B}$. Note that an upper limit set at 10 times the Eddington accretion rate ($\dot{M}_{\rm edd}$) avoids too large super-Eddington ratios during accretion episodes. 

\subsubsection{Galaxy gas disc radius}

In CAM25, which in-turn uses the GAEA SAM, the gas distribution in the galaxy is assumed to have an exponential surface density profile given by:
\begin{equation}
    \Sigma_\mathrm{disc}=\Sigma_0 \exp{\left(-\frac{r}{R_d}\right)}
\end{equation}
where $\Sigma_0=M/2\pi R_d^2$ \citep{Mo98,Hirschmann16,Xie17}. $M$ is the cold gas mass of the galaxy, and $R_d$ is the scale length of the disc. Following the model described in \citet{Guo11}, for a galaxy with non-negligible self-gravity present in an isothermal DM halo, the scale length assuming flat circular velocity curve can be described by:
\begin{equation}
    R_d=\frac{j_\mathrm{gas}}{2V_\mathrm{max}}
\end{equation}
where $j_\mathrm{gas}$ is the specific angular momentum of the cold gas, and $V_\mathrm{max}$ is the maximum rotational velocity of the parent halo \citep{Hirschmann16,Xie17,Zoldan19}. Once the scalelength is calculated for the galaxy, its radius $r_g$ is defined as: 
\begin{equation}
    r_g=3R_d.
\end{equation}

%--------------------------------------------------------------------
\section{Methods}
\label{sec:methods}

\subsection{Seeding scheme}
\label{sec:seed_scheme}

In this work, we adopt a seeding scheme based on light seeds, as our focus includes both SMBHs and IMBHs. We use the implementation from CAM25 (termed as the All Light Seeds, or ALS, scheme in the paper), which in turn uses the scheme described in detail in \citet{Xie17}. According to this scheme, each DM halo is seeded with a BH, with its mass determined based on the mass of the halo ($M_{\rm DM}$) resolved in the simulation, following the relation:
\begin{equation}\label{eq:volonteri_seed}
    M_\text{BH} = \left(\frac{M_\text{DM}}{10^{10} M_{\odot} h^{-1}}\right)^{1.33} \frac{10^{10} M_{\odot} h^{-1}}{ 3 \times 10^{6}}.
\end{equation}
As explained in \citet{Xie17}, this relation is obtained by combining the relations $M_\text{BH}\propto V_c^4$ \citep{Volonteri11}, where $V_c$ is the circular velocity, with $V_c\propto M_\text{DM}^{1/3}$ \citep[][without considering the redshift dependence]{Mo02}. The normalisation is set to reproduce the BH mass - stellar mass relation at $z=0$ \citep{Xie17}. Since we do not have the information of individual stars in a galaxy, and are only using the information of the DM halo to seed the BHs, we do not treat the formation of IMBHs or SMBHs through the runaway stellar or BH merger channels as mentioned in Section \ref{sec:intro}. This means that the BH population generated here is a subset of the entire population and the merger rates of MBHs presented later could be treated as lower limits.

In our cosmological simulation, DM halos are resolved when they reach a mass of $7.9\times10^7 M_\odot$. This implies that this halo will be seeded with a BH of seed mass $4.67\:M_\odot$ by using equation \ref{eq:volonteri_seed}. However, the seeding is done with the CAM25 model, which is based on the GAEA SAM. This SAM is run on a set of merger trees sampled in time as customary in a $N$-body simulation, as if halos were recovered from a set of simulation snapshots. This is in contrast to \textsc{pinocchio}, which outputs a continuous-time merger history of every tree. Thus, in order to use CAM25, the merger history of \textsc{pinocchio} is sampled at various time steps to create snapshots and these are then used as input in CAM25. At earlier epochs, snapshots are sparse and spaced apart in redshift. This implies that a halo which has been resolved by \textsc{pinocchio} at a redshift between two consecutive snapshots would not have its minimum mass when it first appears in the following snapshot due to the growth in mass of this halo via accretion of DM particles.
This higher mass is used in equation \ref{eq:volonteri_seed} to calculate the mass of the BH seed. Consequently, the halo mass used for computing the seed BH mass ranges from the minimum halo mass to $\sim1.85\times10^9 M_\odot$, corresponding to a BH seed mass in the range of $[4.67,311] M_\odot$. This scheme effectively implements the Pop III seeding mechanism.

In our seeding scheme, we also consider a cutoff redshift for seeding the halos. Since Pop III stars require pristine metal-free gas, the early Universe had ideal conditions for their formation. However, due to the feedback from formation and collapse of these stars, further Pop III star formation is reduced or cutoff \citep{Bromm03b,Whalen08a,Whalen08b}. Although it is possible that they are still forming in pockets of pristine metal free gas in the local universe, the star formation rate density is lower compared to the pre-reionization era. Due to the lack of observational constraints, the exact redshift range of their formation is highly model dependent \citep[e.g.][]{Wise12, Johnson13, Mebane18, Jaacks19, Liu20}. In particular, \citet{Mebane18} find that in some models the formation ends at redshifts higher than 10, while in other models they can still form until $z\sim6$ with rates of $10^{-4}$ to $10^{-5}$ $M_\odot$ yr$^{-1}$Mpc$^{-3}$. In the most optimistic models by \citet{Liu20}, who explore their formation below $z\sim6$, it is found that these stars still form at $z\sim0$, although at a low rate of $\sim10^{-7}$ $M_\odot$ yr$^{-1}$Mpc$^{-3}$. In their most pessimistic model, the formation is already terminated by $z\sim5$. We place a lower limit on the seeding redshift at $z=8$, in agreement with the instantaneous reionization redshift measured by Planck Collaboration (\citeyear{Planck20}). This implies that we only follow the evolution and mergers of all the halos which appear (or equivalently get seeded) by $z=8$. We also test this redshift cutoff and find that reducing the redshift does not affect our main results for binaries in the intermediate and supermassive range.

\subsection{Merging delays}
\label{sec:merg_delay}

During a halo merger, the more massive (primary) halo remains in the merger history of \textsc{pinocchio} and the less massive (secondary) (sub)halo disappears from it. In our model, we assume that the primary (secondary) halo hosts the primary (secondary) galaxy at its centre, which in turn hosts the primary (secondary) BH at its centre as well. In order to calculate the merger rate of BHs, we need to derive their merging redshifts. We calculate this by dividing the total time elapsed from halo mergers to BH mergers at their centres into two distinct phases. The first is the time taken from the halo to the galaxy merger, $\tau_{{\rm H}\rightarrow{\rm G}}$. This represents the delay between the merger of the DM halos and the time taken by the secondary galaxy to reach the centre and merge with the primary galaxy. The second phase is the time from the galaxy merger to the BH merger, $\tau_{\rm merge}$, which accounts for additional delay caused by dynamical processes within the merged galaxy. We discuss both of these timescales below.

\begin{figure}
    \centering
    \includegraphics[width=\linewidth]{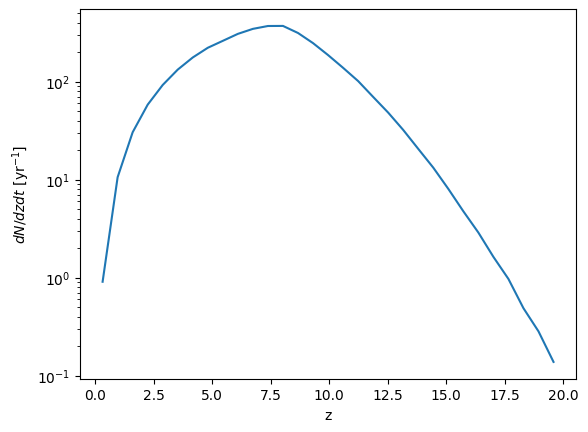}
    \caption{Merging rate of seeded halos (per unit redshift per year) below redshift $z\leq19.92$. The time is shown in observer frame units. The highest redshift corresponds to the highest available snapshot from the CAM25 model; we only consider mergers below this redshift.}
    \label{fig:halo_mergers}
\end{figure}

{\bf $\tau_{{\rm H}\rightarrow{\rm G}}$ -} From \textsc{pinocchio}, we obtain the exact redshift of halo mergers due to its continuous time sampling of the merger history. Using a 40 Mpc/h side length simulation box, we show in Figure \ref{fig:halo_mergers} the halo merger rate for all seeded halos considered in our analysis. To calculate $\tau_{{\rm H}\rightarrow{\rm G}}$ for the mergers, we use the calibrated prescriptions as in CAM25.  
More specifically, this timescale is computed by summing two timescales: 1. The halo survival time, defined as the time taken by the secondary halo since the merger (or equivalently time since accreted onto the main halo) to be stripped to the point that it is not recognised as a distinct sub-halo substructure anymore, which is based on equation 2.6 from \cite{Berner22}; 2. The merging time of the orphan galaxy (the secondary galaxy, hosted in the merged sub-halo), defined as the time taken since the sub-halo DM substructure becomes completely stripped (previous timescale) until the time the galaxy merges with the primary galaxy, which is based on the Chandrasekhar dynamical friction (DF) timescale \citep{Chandrasekhar43} adapted and applied to SAMs \citep[e.g.][]{Boylan-Kolchin08}. 
This prescription requires the halo mass ratio as an input, while the orbital eccentricity is sampled from probability distributions obtained from numerical simulations \citep{Zenter05, Birrer14}. The formulae are calibrated to match the orphan merging time distribution and the stellar mass function at $z=0$ (for more details, refer to CAM25). 

\begin{figure}
    \centering
    \includegraphics[width=\linewidth]{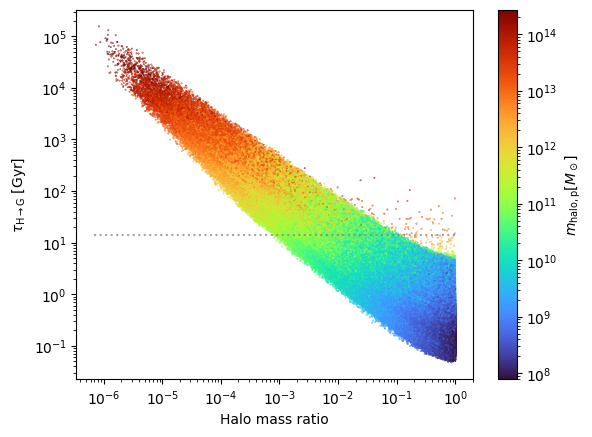}
    \caption{Time taken from halo merger to galaxy merger, $\tau_{{\rm H}\rightarrow{\rm G}}$, as a function of the halo mass ratio. The colour of each point represents the mass of the primary halo. The grey dashed line depicts the Hubble time.}
    \label{fig:halo_to_galaxy}
\end{figure}

Figure \ref{fig:halo_to_galaxy} shows the timescale $\tau_{{\rm H}\rightarrow{\rm G}}$ for the set of halos discussed above and shown in Figure \ref{fig:halo_mergers}. It shows an expected trend \citep[see, e.g.][]{Somerville15}, with mergers involving halos with lower mass ratios taking much longer than the Hubble time to complete. Conversely, halos with mass ratios closer to one are able to complete the galaxy merger within a Hubble time. Although most of these halos have lower mass, there are still a few heavier halos with near-equal masses which merge relatively quickly. The dispersion in the time scale is mainly due to the sampling of orbital eccentricities as mentioned above (CAM25).

{\bf $\tau_{\rm merge}$ -} Once galaxies have merged, we assume that the two BHs are at a separation equal to the radius of the remnant galaxy, which is of the order of a few kpc. Calculating the remaining time taken by the binary to merge is quite uncertain. This involves modelling the internal dynamics of a galaxy and the interaction of stellar objects with the BHs. At separation of $\sim$ kpc, DF is enough to bring the BHs toward the galactic centre. Afterwards, the shrinking takes place mostly due to three-body interactions with stars, which transfer angular momentum away from the binaries into the surrounding stellar population \citep[``stellar hardening''; e.g.][]{Quinlan96,Sesana06,Kelley17,Barber24}. It may happen that the binaries do not undergo enough interactions to shrink further within Hubble timescales. This prevents the BHs from reaching the separations at which the GWs dominate the orbit decay. This issue is called the final parsec problem \citep[e.g.][]{Milosavljevic01,Merritt05}. However, various mechanisms have been proposed to address this problem, such as considering the triaxiality of the galactic bulge, or systems with multiple MBHs, among others \citep[e.g.][]{Yu02,Khan11,Holley15,Sesana15,Bonetti16}. Hence, it is reasonable to assume that the problem is solved. Computing the hardening timescale requires a precise modelling of the galaxy properties, which is out of scope of this work. We neglect this timescale for the simplicity of the model. We further neglect the GW-dominated orbital decay timescale. We only consider the DF timescale for $\tau_{\rm merge}$, since it is the most significant contributor to the delay \citep[e.g. see figure 11 in][]{Langen25}.

To calculate the DF timescale, we use the standard Chandrasekhar prescription \citep{Chandrasekhar43, Binney08}:
\begin{equation} \label{eq:t_dyn}
    T_{\mathrm{dyn}}=19 \left(\frac{r_0}{5\:\mathrm{kpc}}\right)^2\left(\frac{\sigma}{200 \mathrm{~km} / \mathrm{s}}\right)\left(\frac{10^8 \mathrm{M}_{\odot}}{\mathrm{M}_{\mathrm{BH,s}}}\right) \frac{1}{\Lambda}[\mathrm{Gyr}],
\end{equation}
where $r_0$ is the initial separation of the BHs (which we set to the radius of the remnant galaxy), $\sigma$ is the velocity dispersion of the remnant galaxy, and $M_\mathrm{BH,s}$ is the mass of the secondary BH. $\Lambda$ is the Coulomb logarithm given by $\Lambda=\ln{(1+M_\star/M_\mathrm{BH,s})}$, where $M_\star$ is the stellar mass of the remnant galaxy. To estimate the velocity dispersion, we use the scaling relation based on observations \citep{Graham13, McConnell13, Kormendy13}: 
\begin{equation} \label{eq:sigma}
    \frac{M_\mathrm{BH,p}}{10^9 M_\odot} = \left(0.310^{+0.037}_{-0.033}\right)\left(\frac{\sigma}{200 \mathrm{km s}^{-1}}\right)^{4.38\pm0.29}.
\end{equation}
We use only the mean value of the relation since the errors are small.

The Chandrasekhar formula (equation \ref{eq:t_dyn}) assumes that the infalling BH coming from the secondary galaxy is completely stripped of all the stellar component surrounding it. However, many simulations have pointed out that the stripping depends on the morphology of the galaxies, the eccentricity of the orbit, mass ratio, incidence angle, and more \citep[see, e.g.][]{Chang13,Varisco24}. This implies that the primary galaxy would not "see" just a point-like source, the BH, spiralling inward, but rather an extended mass distribution consisting of the stellar component of the secondary galaxy, including the BH. Since the DF timescale is approximately inversely proportional to the infalling mass and carries the biggest uncertainty\footnote{$\sigma$ weakly depends on $M_\mathrm{BH,p}$, and $\Lambda$ is equivalent to $\mathcal{O}(1)$. The radius of the galaxy, which is calculated by using a scaling relation based on the DM halo mass (see next section), does not change significantly if we use the mass of the primary DM halo, or the sum of primary and secondary DM halo masses (see Figure~\ref{fig:bh_int_sep}).} for this timescale, the additional matter surrounding the secondary BH would lead to a reduced DF timescale. This implies that the timescale calculated from equation \ref{eq:t_dyn} can be considered as an upper limit. Since accurately modelling the stripping of the secondary galaxy is beyond the scope of this work, we modify equation \ref{eq:t_dyn} to provide a simple lower-limit estimate of the timescale; we assume that the entire stellar mass of the secondary galaxy surrounds the infalling BH for the whole DF phase:
\begin{equation} \label{eq:t_dyn_ll}
    T_{\mathrm{dyn}}^\mathrm{ll}=19 \left(\frac{r_{0,\mathrm{p}}}{5\mathrm{~kpc}}\right)^2\left(\frac{\sigma}{200\mathrm{~km} / \mathrm{s}}\right)\left(\frac{10^8 \mathrm{M}_{\odot}}{\mathrm{M}_{\star,\mathrm{s}}}\right) \frac{1}{\Lambda^\mathrm{ll}}[\mathrm{Gyr}],
\end{equation}
where the superscript `ll' denotes `lower limit', subscripts `p' and `s' denote the properties of the primary and secondary galaxy respectively, and $\Lambda^\mathrm{ll}=\ln{(1+M_{\star,\mathrm{p}}/M_{\star,\mathrm{s}})}$.

Another assumption used by the Chandrasekhar formula is that the mass of the infalling object must be much larger than the mass of the field stars. Considering a typical initial mass function (IMF) of stars such as the standard four-part power law \citep{Kroupa01, Kroupa02, Weidner04}, and the composite galactic-field IMF obtained by summing over all the stars contained in the clusters within a galaxy \citep{Kroupa03}, most of the stars are less massive than $10\:M_\odot$. Quantitatively, if we consider the galactic-field IMF from \cite{Kroupa03} (figure 1 of the paper) with the mass of stars ranging from 0.01 $M_\odot$ to 150 $M_\odot$, then more than 99.92\% of the stars are less massive than $10\:M_\odot$ within a typical galaxy. This implies that the DF timescales calculated would be valid for almost all the inspiralling BHs more massive than $\sim10\:M_\odot$. Furthermore, in this work, we focus solely on estimating the merger rate for BHs in the intermediate and supermassive range which could be detected by LGWA and LISA. With these considerations, we only consider BHs more massive than $10\:M_\odot$ in our analysis.

As evident from the equations reported, the calculation of the DF requires the knowledge of the galaxy properties such as the mass and radius, as well as the mass of the BHs. To compute these quantities, we use the CAM25 model. In the next section, we explain how we use the outputs of the model for calculating these properties.

\subsection{Scaling relations}
\label{sec:scaling_rel}

\begin{figure*}
    \centering
    \includegraphics[width=\linewidth]{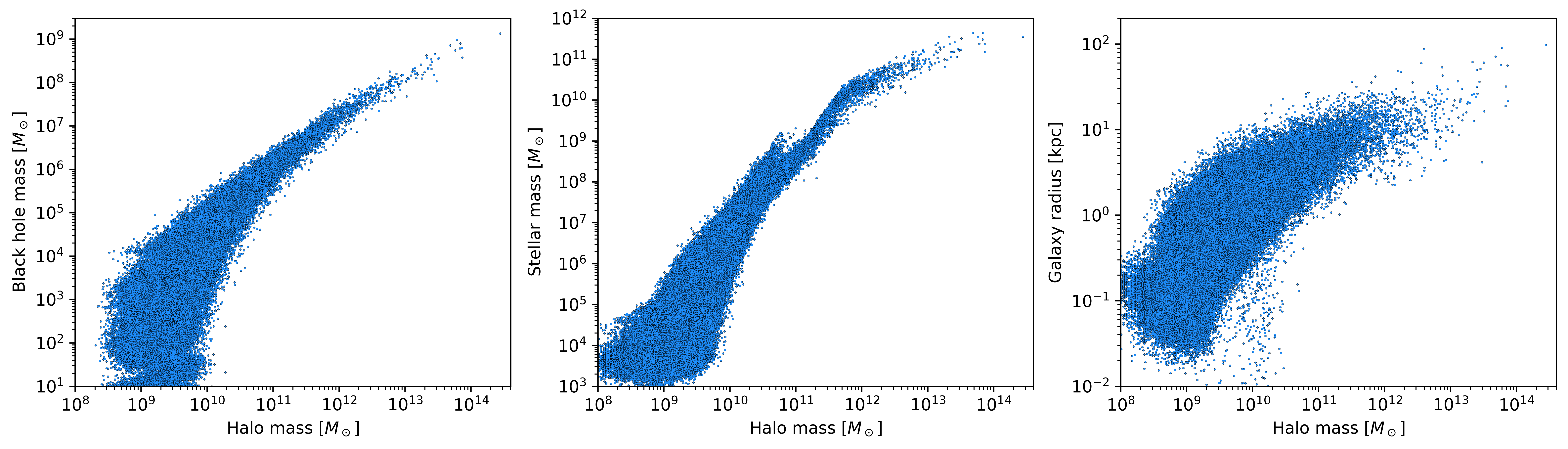}
    \caption{The BH mass, galaxy mass, and the galaxy radius as a function of halo mass from the CAM25 model with ALS seeding at $z=0$.}
    \label{fig:pingaea_data}
\end{figure*}

As mentioned in Section \ref{sec:seed_scheme}, the CAM25 model provides the baryonic properties, including the BH mass $M_\mathrm{BH}$, galaxy stellar mass $m_*$, and the radius $r_\mathrm{g}$ at discrete redshift snapshots. In Figure \ref{fig:pingaea_data} we present respectively, their scatter plots with respect to the halo mass at $z=0$. The merger history of \textsc{pinocchio}, instead, has continuous time sampling, which means that it provides the exact redshift of each halo merger. Since it is a DM-only code, the output also consists of the exact halo mass at the merging redshift. This motivates us to find scaling relations of the baryonic properties with respect to the DM halo mass. These relations  allow us to use the halo mass at the exact merging redshift to calculate all the required properties used to evaluate the delays. Below, we describe the method we used to create the scaling relations.

From each available CAM25 output at a redshift $z_i$, we divide the entire halo mass range into small bins $[m_1,m_2]$ and create probability density functions (PDFs) $f(\Theta|z_i, m_1, m_2)$ for each property $\Theta\in\{M_\mathrm{BH}, m_*, r_\mathrm{g}\}$ separately, for all the halo mass bins. This allows us to construct a framework where all the baryonic properties of interest are dependent only on the redshift and the halo mass. To construct these PDFs, we generate histograms of the quantities $\Theta$ within small halo mass bins\footnote{The halo mass bins are small enough to preserve the correlation among all the properties.} and use the \texttt{scipy.stats} library to convert the histograms to PDFs. 

To understand how we use these PDFs, let us say that we need to evaluate a baryonic property $\Theta$ of a halo with mass $m_\mathrm{halo}$, at a redshift $z$. To calculate these properties, we use the PDFs of that property that we constructed above at the redshifts $z_1$ and $z_2$ where the CAM25 outputs are available, such that $z\in[z_1,z_2]$, both for the same halo mass bin $[m_1,m_2]$ where $m_\mathrm{halo}\in[m_1,m_2]$. The two PDFs at those redshifts are $f_i(\Theta)=f(\Theta|z_i,m_1,m_2)$ for $i=1$ or 2. Then we linearly interpolate the PDFs in redshift space to compute the exact PDF at the required redshift $z$ as follows:
\begin{equation}
    f'(\Theta|z,m_1,m_2)=\left(\frac{z_2-z}{z_2-z_1}\right)f_1(\Theta) + \left(\frac{z-z_1}{z_2-z_1}\right)f_2(\Theta),
\end{equation}
where $f_i(\Theta)=f(\Theta|z_i,m_1,m_2)$. The redshift intervals of the CAM25 outputs are small enough such that the linear interpolation is sufficient in preserving the evolution of the baryonic properties. Once we calculate this distribution $f'$, we compute the required PDF by normalising it:
\begin{equation}
    f(\Theta|z,m_1,m_2) = \frac{f'(\Theta|z,m_1,m_2)}{\int f'(\Theta|z,m_1,m_2) d\Theta},
\end{equation}
where the integral in the denominator is performed over the minimum and maximum values of $\Theta$ allowed by the model. This method allows us to evaluate the baryonic properties at the exact merger redshifts using the DM halo mass provided by the merger history of \textsc{pinocchio}.

\subsection{Calculated delays}
\label{sec:merg_pop}

Using all the relations mentioned above, we finally calculate all the properties required to evaluate the DF timescale for all the merging galaxies and BHs. In the subsequent analysis, we only consider those galaxies that either reach a stellar mass of $10^5 M_\odot$ by $z=0$, or merge into a larger galaxy that reaches this mass by $z=0$. This approach avoids imposing a fixed galaxy mass cut across all redshifts, which would be inappropriate given that galaxies at higher redshifts are less massive. The galaxy mass condition, along with the BH mass threshold of $10\:M_\odot$ and a formation redshift cutoff of $z=8$, are the only selection cuts used to define the merging population.

\begin{figure}
    \centering
    \includegraphics[width=\linewidth]{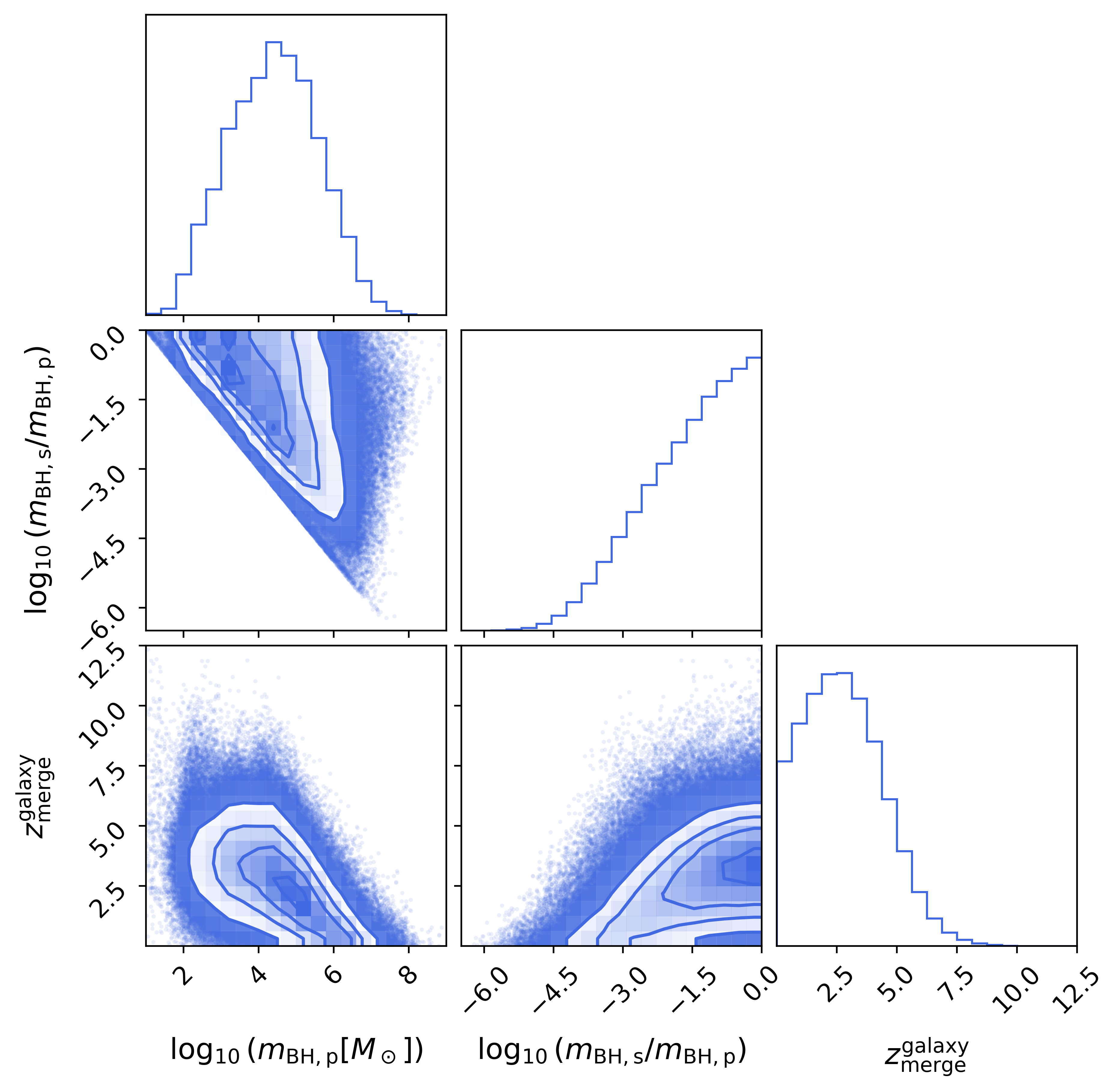}
    \caption{Corner plot showing the distribution of the population of BH pairs formed after the galaxy merger. The plot shows the primary mass $m_\mathrm{BH,p}$, the secondary mass $m_\mathrm{BH,s}$ divided by the primary BH mass $m_\mathrm{BH,s}/m_\mathrm{BH,p}$, and the redshift of the galaxy merger $z_\mathrm{merge}^\mathrm{galaxy}$. The contours show the density of the points. Individual points outside the contours are also shown to visualize the extent of the distributions.}
    \label{fig:bh_mass_cut}
\end{figure}

Once we use the scaling relations and apply all the aforementioned cuts, we obtain the population of BH pairs shown in Figure~\ref{fig:bh_mass_cut}, for which we calculated the DF timescales for the inspiral. These are the pairs that form after adding the time delay $\tau_{{\rm H}\rightarrow{\rm G}}$ (Figure \ref{fig:halo_to_galaxy}) to the halo merger redshifts. Although Figure \ref{fig:halo_to_galaxy} shows that only halos with high mass ratios merge, it is evident from Figure \ref{fig:bh_mass_cut} that BH pairs spanning mass ratios of more than six orders of magnitude form within the age of the universe. This is due to a combination of the growth of BH and galaxies in the CAM25 model, which results in the larger variance on the dependence of these masses at lower halo masses, and most of these halos are the ones that end up merging within a Hubble time. 

\begin{figure}
    \centering
    \includegraphics[width=\linewidth]{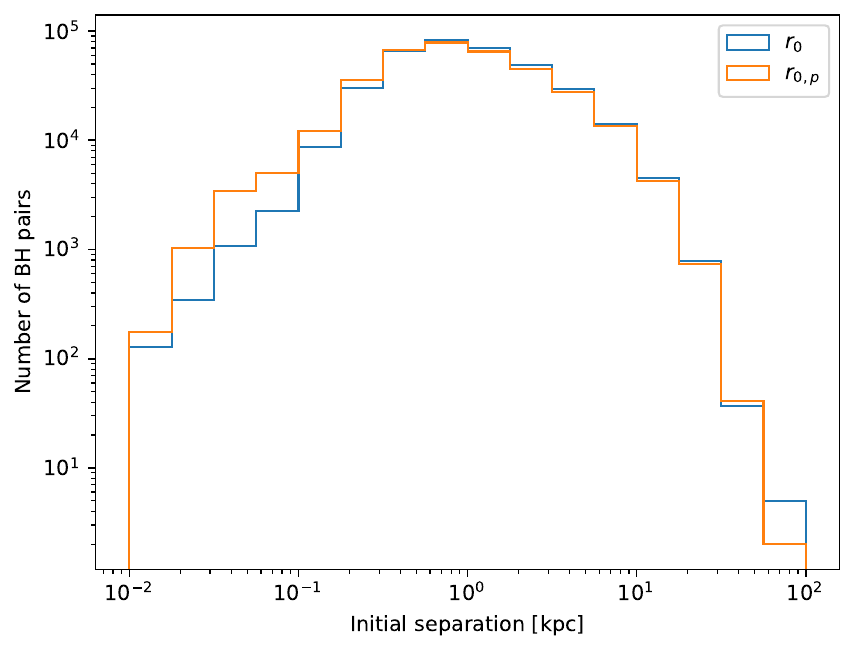}
    \caption{Histogram of the initial BH separations used for computing the DF timescale. $r_0$ corresponds to the separation used to evaluate the upper limits on the timescale (equation \ref{eq:t_dyn}), while $r_{0,p}$ is used for the lower limits (equation \ref{eq:t_dyn_ll}).}
    \label{fig:bh_int_sep}
\end{figure}

As discussed in Section~\ref{sec:merg_delay}, after the galaxy merger, BHs are assumed to have a separation equal to the radius of the remnant galaxy, $r_0$, in the upper limit case (equation \ref{eq:t_dyn}), while in the lower limit case, separation equal to the radius of the primary galaxy $r_{0,p}$ is assumed. Figure~\ref{fig:bh_int_sep} shows a histogram of separations for both these cases. The distributions are quite similar, with most pairs having separations of the order of one kiloparsec. The difference is mostly due to small excess of BHs at smaller separations for the $r_{0,p}$ case. This is because we use the mass of the primary halo to calculate the radius of the primary galaxy, which is smaller than the mass of the remnant halo used to calculate the radius of the remnant galaxy (see Section \ref{sec:scaling_rel}). Due to the small difference in their respective halo masses, the resultant radii calculated are similar. 

\begin{figure*}
\sidecaption
    \centering
    \includegraphics[width=12cm]{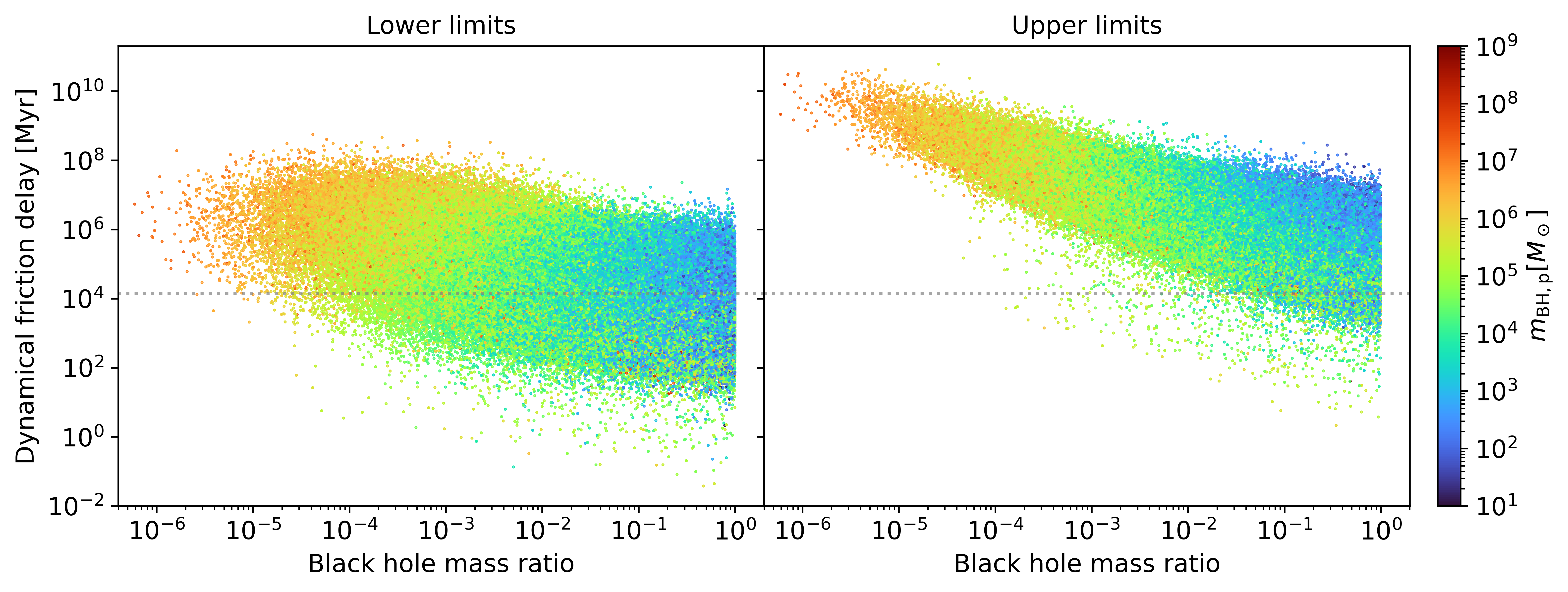}
    \caption{The lower and upper limits of DF delays in Myr computed using equations \ref{eq:t_dyn_ll} and \ref{eq:t_dyn}, respectively. The colour represents the mass of the primary BH. The dashed horizontal line shows the age of the universe.}
    \label{fig:dyn_fric_del}
\end{figure*}

Next, we finally calculate both the upper and lower limits of the DF timescale using equations \ref{eq:t_dyn} and \ref{eq:t_dyn_ll} and present the results in Figure \ref{fig:dyn_fric_del}. We can see that the high BH mass ratio pairs have a higher tendency to merge within a Hubble time, with a lower scatter in the upper limits case. This can be explained by considering equation \ref{eq:t_dyn} used for evaluating the timescale. If we ignore the dependence on $\sigma$ and $\Lambda$ based on arguments presented in section \ref{sec:merg_delay}, then $T_\mathrm{dyn}\propto r_0^2/M_\mathrm{BH,s}$. Since more massive galaxies, especially those with larger halos (see Figure \ref{fig:pingaea_data}), tend to have larger central BHs and radii, the radius dependence can be approximated by the primary BH’s mass, implying: $T_\mathrm{dyn}\propto M_\mathrm{BH,p}^2/M_\mathrm{BH,s}=(M_\mathrm{BH,p}/$BH mass ratio). This approximation shows that the timescale is higher for pairs with high primary BH masses and low mass ratios, while it is lower for pairs with lower primary masses and higher mass ratios. We note that these approximations are comparatively more valid for larger halo masses because there is a higher spread in the baryonic properties at lower halo masses. This weakens the correlations and results in the spread of the timescales, as seen in Figure \ref{fig:dyn_fric_del}. In the lower limits case, the dependence on the mass ratio is weaker because of the fact that we used the mass of the secondary galaxy instead of the secondary BH to evaluate the delay. This results in a slightly weaker dependence on the BH mass ratio (only slightly because there is still a correlation of the BH and galaxy mass, although weaker at these halo masses).

\begin{figure}
    \centering
    \includegraphics[width=\linewidth]{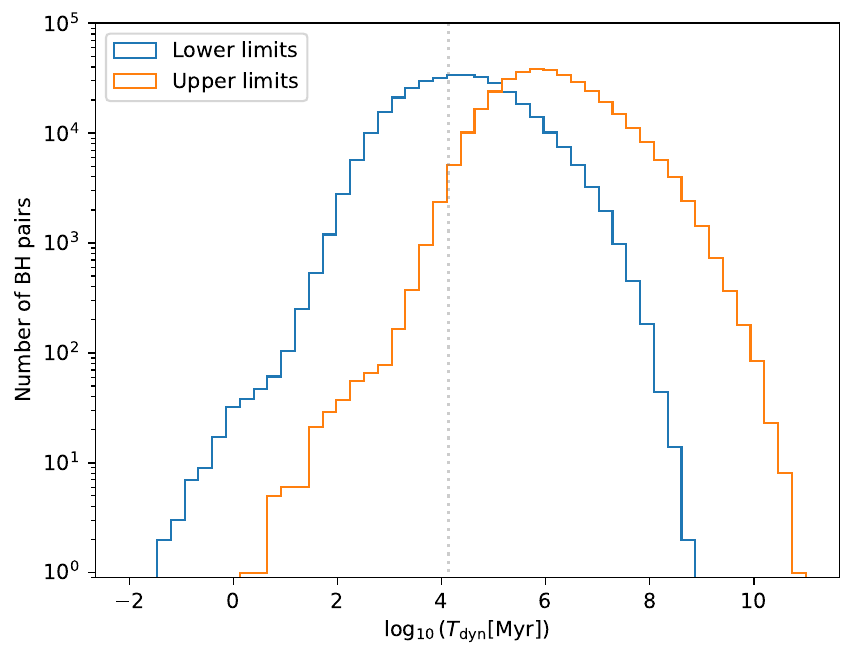}
    \caption{Histogram of the DF delays for all the pairs of BHs. The dashed line marks the age of the universe. The lower and upper limits corresponds to the rate calculated using equations \ref{eq:t_dyn_ll} and \ref{eq:t_dyn}, respectively.}
    \label{fig:t_dyn_hist}
\end{figure}

As a result of these trends, less than 2\% of binaries have the DF delay less than the Hubble time in the upper limits case, while around 40\% binaries in the lower limits case. This can be seen from the histogram of the timescales presented in Figure \ref{fig:t_dyn_hist}. Furthermore, among these binaries, both the cases have around $\sim$60\% of the binaries with total mass in the IMBH range and $\sim$30\% in the SMBH range. In the next section, we discuss the rate of mergers for all these BH pairs.

\section{Results}
\label{sec:results}

\subsection{Merging population}

\begin{figure*}
\sidecaption
    \centering
    \includegraphics[width=12cm]{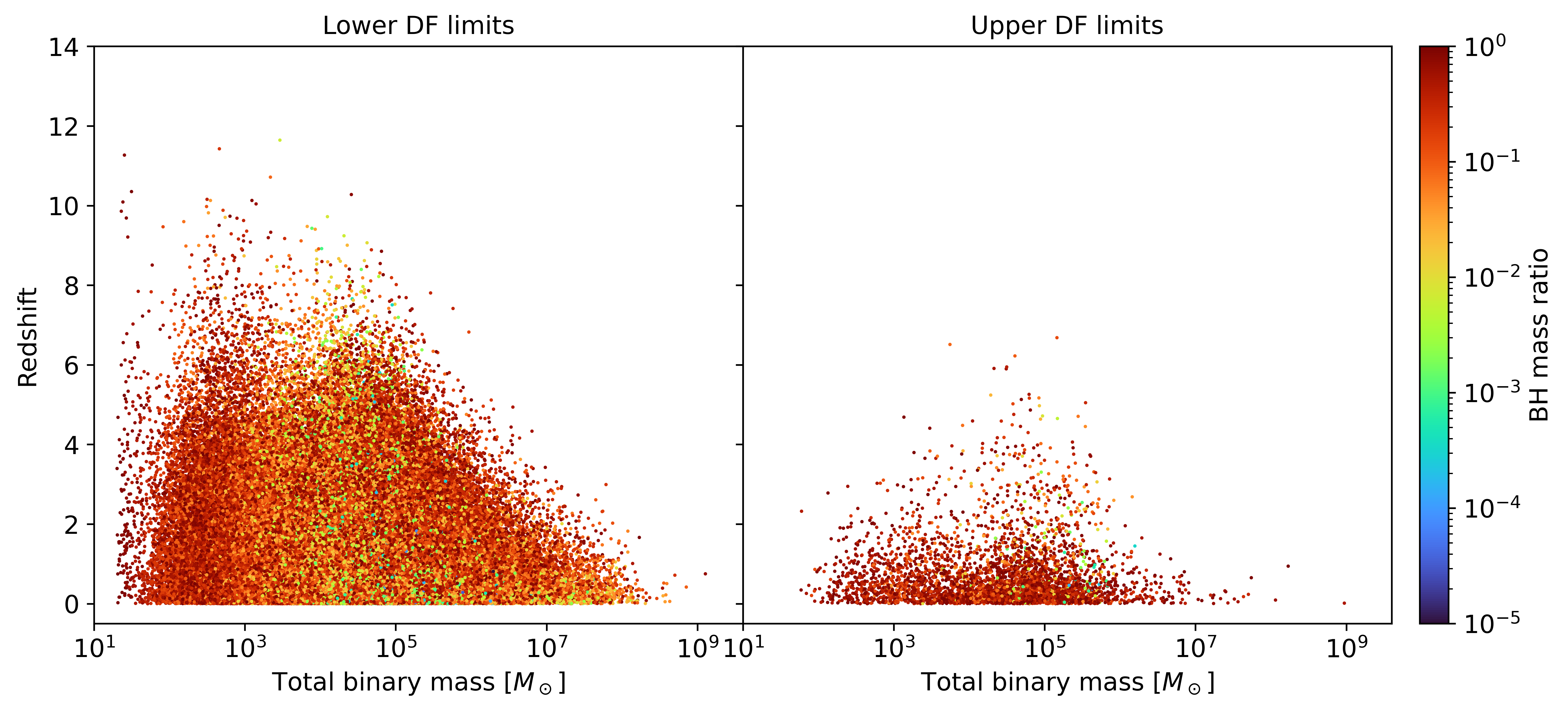}
    \caption{The merging redshifts of all the BH pairs, with respect to the total binary mass for both the limits of DF timescale. The colour bar represents the binary mass ratios.}
    \label{fig:merging_bhs}
\end{figure*}

After applying the DF delays and assuming that the BHs merge soon after (see Sect. \ref{sec:merg_delay}), we generate a population of merging BHs for both the lower and upper limits of the DF timescale. Figure \ref{fig:merging_bhs} shows all these merging BH pairs along with their corresponding merger redshifts. There is a significant difference in the number of BH pairs merging for the two cases, consistent with the discussion in section \ref{sec:merg_pop}. 

The median redshift of mergers for the lower limit case is 1.18, while it is even lower for the upper limits case at 0.45. This indicates that most of the mergers occur at low redshifts. Only a handful of binaries merge above $z=10$ in the lower limits case, while none merges for the upper limits. Although we are considering all the halo mergers below $z\approx20$, the actual rate of merger at such high redshifts is quite low. The time span from $z=20$ to $z=10$ is only around $\sim300$ Myr, which is comparable to the minimum values of the additional delays that we applied, thereby limiting the number of mergers that can take place at high redshift.

Most merging pairs have mass ratios larger than 0.01: over $\sim$90\% of the total merging population for the lower DF limits, and more than 98\% for the upper DF limits. For the mass range of MBHs we are considering, 87\% of binaries with a total mass in the IMBH regime have mass ratios higher than 0.01 in the lower DF limits case, increasing to over 98\% for the upper DF limits case. A similar trend holds for the SMBH regime, with more than 87\% of binaries mergers with mass ratio higher than 0.01 in the lower-limit case, and more than 95\% in the upper-limit case. These results imply that most mergers are major, with only a very small percentage intermediate and extreme mass ratio inspirals and mergers.

\begin{figure*}
\sidecaption
    \centering
    \includegraphics[width=12cm]{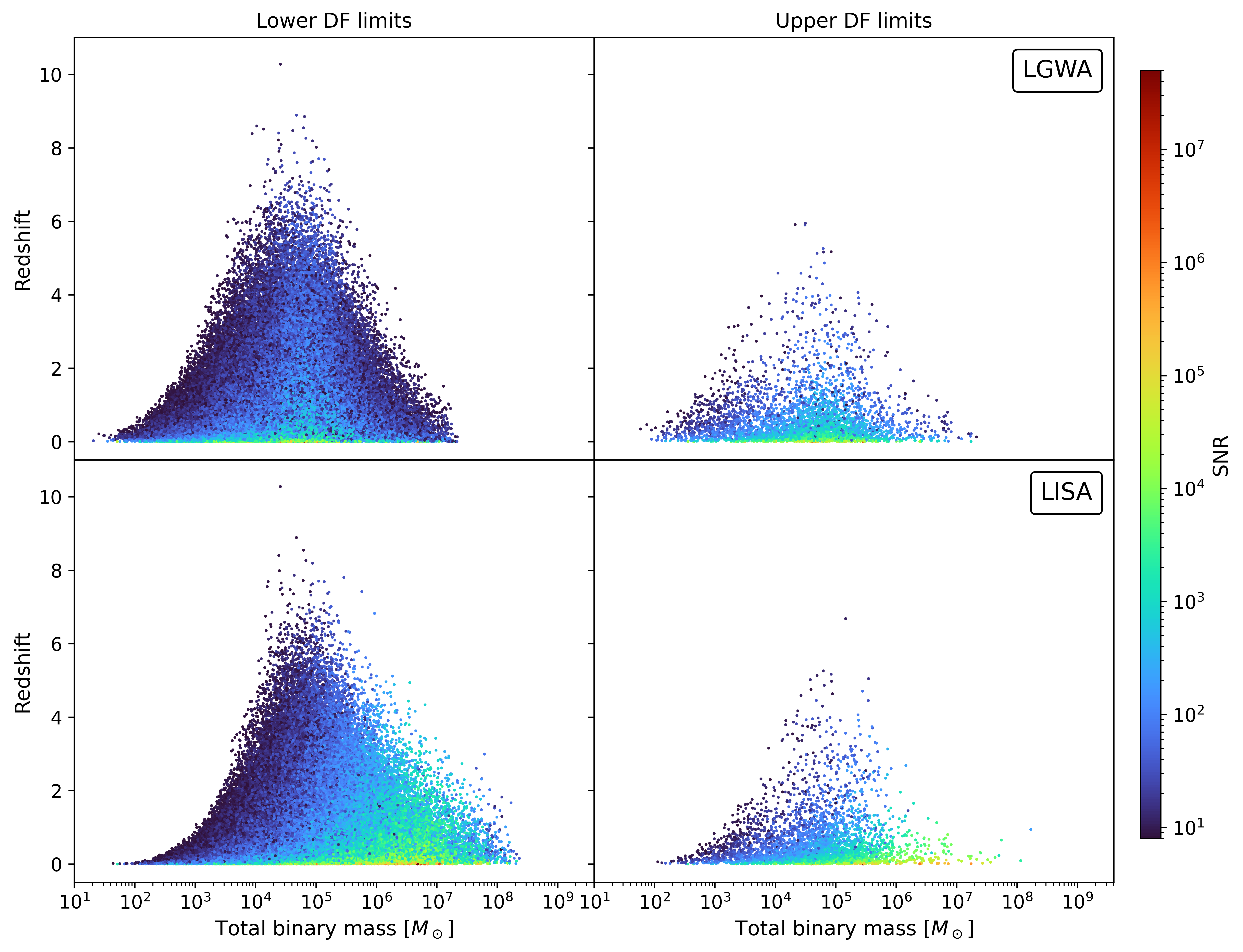}
    \caption{The SNR of all the merging events as seen from LGWA (top row) and LISA (bottom row). Only the events with SNR>8 are shown.}
    \label{fig:snr_all}
\end{figure*}

In order to calculate the detectability of these mergers by GW detectors, we assign random position in the sky for each source and compute the SNR using the \texttt{GWFish}\footnote{\url{https://github.com/janosch314/GWFish}} library \citep{Dupletsa23}. The SNRs are calculated using the \texttt{IMRPhenomD} waveform approximant \citep{Husa16,Khan16} from the LALSimulation \citep{lalsuite}, and by assuming observation durations of 10 years for LGWA and 4 years for LISA. Figure \ref{fig:snr_all} presents the SNR of events detectable by LISA and LGWA under both the lower and upper DF timescale limits. To keep our analysis simple, we assume all the BHs are spinless and they do not receive a kick after the merger which can potentially eject them from the galaxy centre \citep{Campanelli07,Blecha08}. We assume that all events with SNR > 8 are detectable. As expected from the design sensitivities of the detectors, LGWA is able to detect more binaries with intermediate-mass range and with higher SNRs, compared to LISA.

\subsection{Merger rate}

To track the evolution of the rate of mergers, we calculate the quantity $dN/dtdz$ which gives the number of mergers per unit redshift and year by using the relation:
\begin{equation} \label{eq:merg_rate}
    \frac{dN}{dtdz}(z) = \frac{dn}{dz}(z)\times4\pi cd_c(z)^2,
\end{equation}
where $dn/dz$ is the differential number of BH mergers per unit comoving volume per differential redshift bin, and $d_c(z)$ is the comoving distance at redshift $z$. The time is in observer frame units. Due to the statistical nature of our model, we perform our entire analysis twenty times to calculate an average rate with dispersion. The benefit of performing the analysis multiple times helps in building statistics, especially at higher redshifts where the number of mergers is quite low. The merger rates corresponding to the lower DF limit is regarded as the optimistic case, while the rates corresponding to the upper DF timescale limits as the pessimistic case. In the next two subsections, we present the rates considering the mergers of all the BHs, and then for specific mass ranges of interest.

\subsubsection{All masses}

In Figure \ref{fig:merger_rate_all}, we present the rates for all masses of BHs. The left panel shows the rates for all the mergers, without convolving with the sensitivity curves of the two detectors. The merger rate curves follow the shape of the halo merger rate (Figure \ref{fig:halo_mergers}). However, due to the inclusion of delays, the peak of mergers shifts to a lower redshift. In the optimistic case, the peak occurs around $z\sim2$, while in the pessimistic case, it shifts to $z\sim1$. At the high redshift end, mergers start happening at $z\sim14$ for the optimistic case, although the dispersion is large due to both a low number of mergers and the dispersion in merging time delays. For the pessimistic case, mergers occur only below $z\sim8$. Although the difference in the rates from the two cases is large at higher redshifts, it decreases at lower redshifts, reaching a difference of only around an order of magnitude by $z=0$.

In the remaining panels of the figure, we present the rates after convolving with the sensitivity curves of LISA and LGWA. For all the sources with SNR > 8, the rates for both the detectors are similar. For the rates from the pessimistic case, the rates are also quite similar with the rate from all the mergers in the left panel. This implies that most of the merging population constructed by considering the upper limits of DF timescales will be detectable by both detectors. The rates from the optimistic case are slightly lower than the rate for all the mergers, especially at higher redshifts. LGWA is able to detect at slightly higher redshifts of around $\sim11$ compared to LISA. But the overall distribution of rates for both the detectors is quite similar. However, the difference emerges for sources with SNR>100, for the optimistic case rates. LISA is able to detect more mergers, and up to $z\sim7$, while LGWA does not see mergers with such high SNR at redshifts beyond $z\sim5$. The pessimistic case rates is still quite similar.

\subsubsection{Mass ranges}

To analyse the rates for specific BH mass ranges, Figure \ref{fig:merg_rate_mass} shows the evolution of merger rates for three total binary mass intervals - $[10^3,10^5] M_\odot$, $[10^5,10^7] M_\odot$, and $[10^7,10^8] M_\odot$ - each evaluated with two different SNR cuts. The first column also shows the rates for all the merging binaries, without any SNR cut. For the intermediate mass range binaries, the rate of mergers detected by both the detectors with SNR > 8 is quite similar, for both the optimistic and pessimistic cases. LGWA has a slightly higher rate and is able to detect mergers at slightly higher redshifts in the optimistic case, reaching up to $z\sim11$. However, the difference becomes much more pronounced for events detected with SNR higher than 100, for both cases. This is expected since LGWA has a higher sensitivity to these mass ranges compared to LISA.

For more massive BHs ($10^5$ to $10^7 M_\odot$), the roles reverse and LISA has a higher detection rate. Again, the difference in the rates is not large for detections with SNR > 8 for both cases. At a higher SNR cut, the detection rates for these binary BHs by LISA are significantly higher and reach redshifts higher than those for LGWA, especially in the optimistic case. In the pessimistic case, the difference is still present, although less pronounced. This implies that the similar rates observed in the smaller SNR cut are due to the fact that LISA is detecting a similar number of mergers, but with much higher SNR. Finally, for the most massive BHs ($10^7$ to $10^8 M_\odot$), LISA detects significantly more events in the optimistic case and also reaches higher redshifts than LGWA. In the pessimistic case, the rates are similar for the lower SNR cut. Note that the evolution of the rate in the optimistic case for LISA is quite similar for both the SNR cuts, which is in turn similar to the rate of all the mergers in this mass range. This implies the LISA is able to detect most of those mergers with a high SNR.

\subsubsection{Total detections per year}

\begin{figure*}
    \centering
    \includegraphics[scale=0.58]{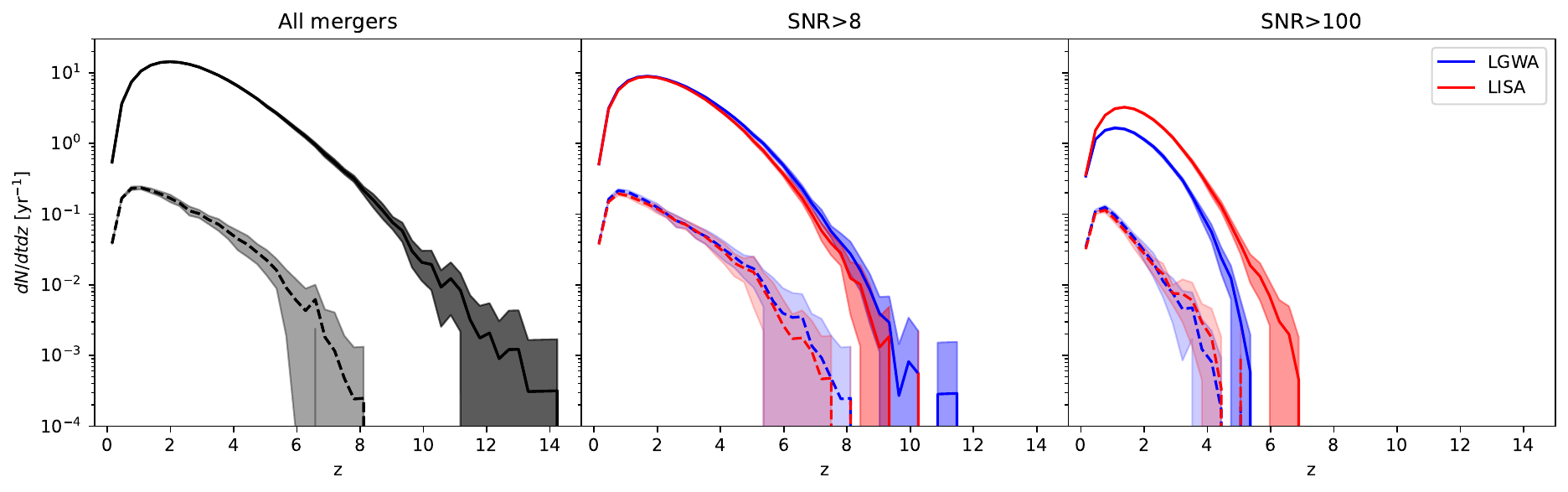}
    \caption{The number of mergers per unit redshift per year at different redshifts detected by the two detectors for two cuts on the SNRs. The time is reported in the observer frame. The lines are the average of the rates over 20 realizations, while the shaded region represents $1\sigma$ deviation from the mean. The solid lines and darker shaded region corresponds to the rate from considering the lower limits on DF, while the dashed lines and lighter shaded region corresponds to the upper limits. The left panel shows the rate for all the mergers in our model, while the middle and right panels shows the rate for mergers detected by LGWA (blue) and LISA (red) with a minimum SNR of 8 and 100 respectively.}
    \label{fig:merger_rate_all}
\end{figure*}

\begin{figure*}
    \centering
    \includegraphics[scale=0.61]{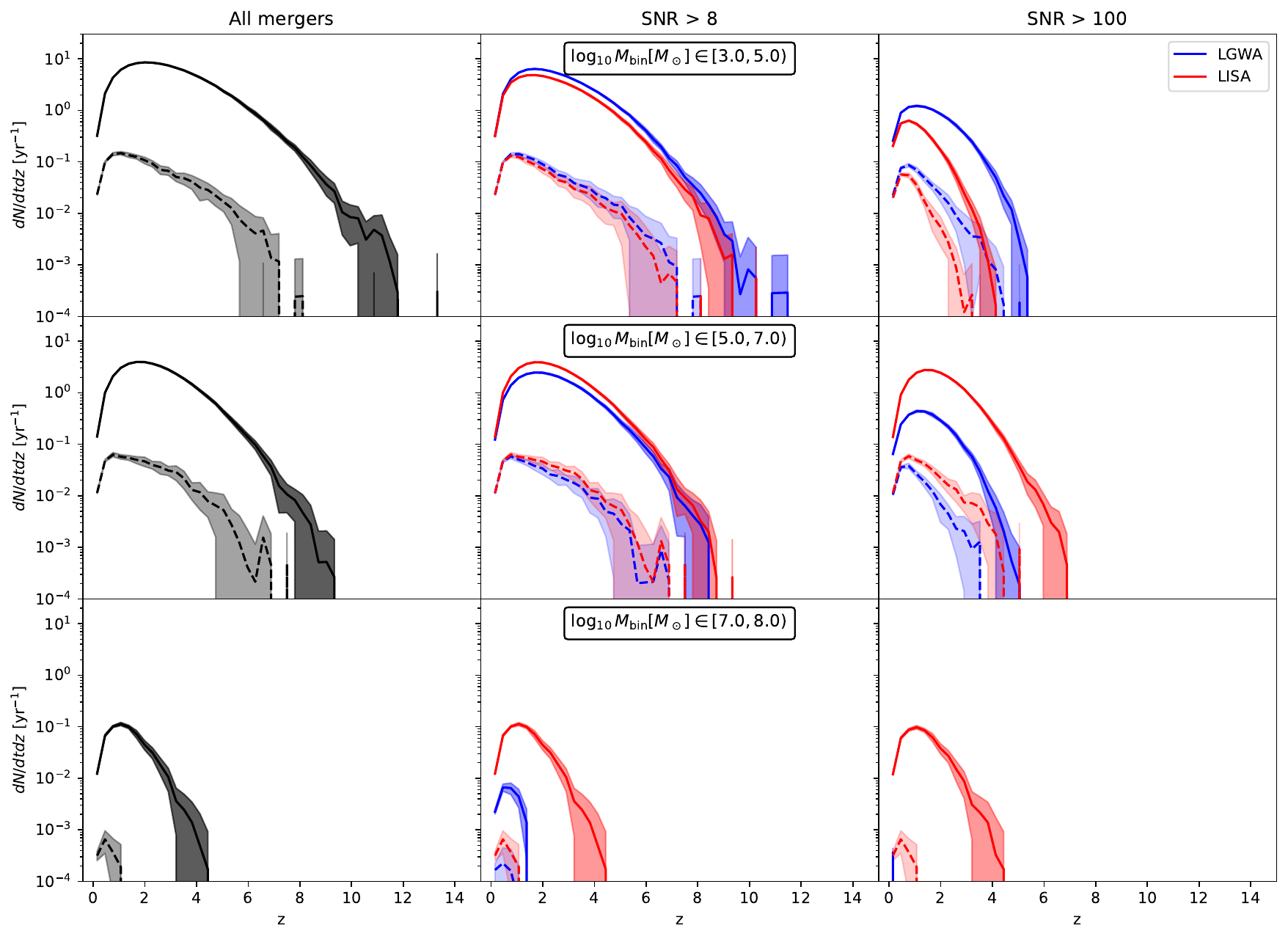}
    \caption{The merger rate evolution in observer frame of all binaries within mass ranges $[10^3,10^5] M_\odot$, $[10^5,10^7] M_\odot$, and $[10^7,10^8] M_\odot$. The first column shows the rate for all the mergers, while the remaining columns show the rate for events with SNR > 8 and 100 as seen by the detectors. The colour scheme is the same as in Figure \ref{fig:merger_rate_all}.}
    \label{fig:merg_rate_mass}
\end{figure*}

To calculate the total number of detections per year, we integrate $dN/dtdz$ obtained from equation \ref{eq:merg_rate} over the entire redshift range. Table \ref{table:merger_rate} reports the total number of mergers per year in the observer frame after the integration. These numbers represent the averages from our twenty realisations of the merging population, as shown in Figures \ref{fig:merger_rate_all} and \ref{fig:merg_rate_mass}. For all the BH binaries, the pessimistic case predicts an average of only 0.61 mergers per year, while in the optimistic case this number rises to nearly 50 per year. Of these, both LGWA and LISA would detect about 0.5 mergers per year in the pessimistic scenario, increasing to more than 26 detections per year in the optimistic one. Despite the significant difference in the number of possible detections among the two cases, with the expected lifetime of LGWA (10 years) and LISA (4 years), even the pessimistic case results in some detections with SNR > 8. This implies that within the lifetime of these detectors, there is a high probability of a handful of detections. 

For the binaries with total mass in the intermediate range, detection numbers are low. However, over their expected lifetime, LGWA should detect at least a few mergers, and LISA may observe at least one in the pessimistic case. In the optimistic scenario, both detectors could observe many mergers. Furthermore, some sources with higher SNR should also be detectable. As expected, LGWA has higher rates of detection compared to LISA for these masses of binaries. For the binaries in the supermassive regime, the detectability tendency flips, and LISA presents the possibility to detect more mergers than LGWA. For binaries in the range from $10^5~M_\odot$ to $10^7~M_\odot$, the difference in the rates in the pessimistic case is low, although the difference in the lifetime of these detectors may play a major role in the total number of detections. In the optimistic case, there are a lot more detections, with at least 7 per year for LGWA and more than 11 per year for LISA. As the case for LGWA for lower mass binaries, LISA will also be able to detect higher mass binaries with a higher SNR. Finally, for the most massive BHs, the rates are low, but expected since these detectors are not expected to be able to detect their mergers within their lifetimes.

The similar detection rates for sources with  SNR > 8 in both detectors suggest a strong potential for synergy. If both of them are active at the same time, the lower frequency sensitivity of LISA allows a potential binary to be seen at an inspiral stage, which would then merge at frequencies sensitive to LGWA. This is especially true for BH binaries with masses around $\sim10^3~M_\odot$ to $10^4~M_\odot$ \citep{Colpi24}. Given the high detection rates for these masses, this synergy offers a valuable opportunity to study a still elusive BH population and place crucial constraints on the BH mass function in this range.

\subsubsection{Comparison with other models}

As mentioned briefly in the introduction, the predicted merger rate from different models depends heavily on the model assumptions and the considered physical processes, in particular the seed BH mass and the merging delays. Furthermore, the rates also depend on the simulation resolution, both spatial and in mass, the size of the volume, the simulated redshift range, and more. Due to all these differences among models, a faithful comparison is not entirely feasible. Nonetheless, estimates from different models can help us in understanding the range of detections to be expected from LISA and LGWA.

To date, most population studies have been done with respect to the observational sensitivity of LISA. For instance, in the semi-analytical formulation of \citet{Ricarte18} with their light seed implementation, they predict that LISA will detect around $\sim 40$ to $\sim245$ mergers per year with SNR larger than 5 in their pessimistic and optimistic models, respectively. These models are based on the merging probability of BHs after the parent halos merger. For the heavy seeds, the rates drop to $\sim5$ to $\sim48$ per year for the two cases.

Conversely, \citet{Volonteri20} presents a lower merger rate of MBHs from the high-resolution simulation \mbox{{\sc \small NewHorizon}} \citep{Dubois21}. Depending on the adopted delays, the rate varies from $\sim1$ to $\sim10$ MBH merger per year. Note that these are total merger rates from the simulation, and are not convolved with the sensitivity curve of LISA. In the work of \citet{Barausse20}, which is based on the SAM first presented in \citet{Barausse12}, the detection rate is $\gtrsim2$ mergers per year in LISA, irrespective of the BH seeding scheme. However, the rates significantly depends on the delays and the impact of supernova feedback on MBH growth.

\citet{Kritos25} implemented the formation and evolution of MBHs in the centre of Nuclear Star Clusters using the galaxy merger trees from \mbox{{\sc \small NewHorizon}} and applying the \mbox{{\sc \small Nuce}} SAM \citep{Kritos24}. They report a merger rate of $\simeq5.3$ per year up to $z=5$. Compared to all these works and the literature, the predictions of our model are consistent. Measurements of the rate from LISA will help in constraining these models, allowing us to better understand the physical processes driving MBH mergers.

\begin{table*}[]
\centering
\caption{The number of mergers detected per year from the two detectors for different SNR cuts, summed over the entire redshift range considered. The first main row corresponds to all the BHs irrespective of the mass, while the remaining three correspond to the total mass of the binary $M_\mathrm{bin}$ as mentioned.}
\begin{tabular}{@{}ccccc@{}}
\toprule
                      & \multicolumn{4}{c}{Merger rate {[}yr$^{-1}${]}}                                                                           \\ \midrule
                      & \multicolumn{4}{c}{All masses}                                                                                            \\
                      & \multicolumn{2}{c}{Pessimistic}                             & \multicolumn{2}{c}{~~~~~~~~~Optimistic}                              \\
                      & LGWA                         & LISA                         & LGWA                         & LISA                         \\ \cmidrule(l){2-5} 
All mergers           & \multicolumn{2}{c}{0.61$\pm$0.07}                           & \multicolumn{2}{c}{~~~~~~~~~49.11$\pm$0.85}                          \\
SNR\textgreater{}8    & 0.49$\pm$0.06                & 0.45$\pm$0.06                & 27.60$\pm$0.53               & 26.20$\pm$0.50               \\
SNR\textgreater{}100  & 0.17$\pm$0.02                & 0.16$\pm$0.02                & 3.52$\pm$0.12                & 7.25$\pm$0.21                \\

\multicolumn{1}{l}{}  & \multicolumn{1}{l}{}         & \multicolumn{1}{l}{}         & \multicolumn{1}{l}{}         & \multicolumn{1}{l}{}         \\
                      & \multicolumn{4}{c}{$M_\mathrm{bin}\in[10^3,10^5] M_\odot$}                                                                \\ \cmidrule(l){2-5} 
All mergers           & \multicolumn{2}{c}{0.39$\pm$0.06}                           & \multicolumn{2}{c}{~~~~~~~~~30.34$\pm$0.69}                          \\
SNR\textgreater{}8    & 0.34$\pm$0.06                & 0.29$\pm$0.05                & 19.89$\pm$0.47               & 14.91$\pm$0.41               \\
SNR\textgreater{}100  & 0.12$\pm$0.02                & 0.06$\pm$0.01                & 2.66$\pm$0.11                & 0.91$\pm$0.05                \\

\multicolumn{1}{l}{}  & \multicolumn{1}{l}{}         & \multicolumn{1}{l}{}         & \multicolumn{1}{l}{}         & \multicolumn{1}{l}{}         \\
                      & \multicolumn{4}{c}{$M_\mathrm{bin}\in[10^5,10^7] M_\odot$}                                                                \\ \cmidrule(l){2-5} 
All mergers           & \multicolumn{2}{c}{0.16$\pm$0.04}                           & \multicolumn{2}{c}{~~~~~~~~~11.34$\pm$0.32}                          \\
SNR\textgreater{}8    & 0.13$\pm$0.03                & 0.16$\pm$0.04                & 7.18$\pm$0.26                & 11.08$\pm$0.30               \\
SNR\textgreater{}100  & 0.05$\pm$0.01                & 0.10$\pm$0.02                & 0.85$\pm$0.06                & 6.18$\pm$0.19                \\

\multicolumn{1}{l}{}  & \multicolumn{1}{l}{}         & \multicolumn{1}{l}{}         & \multicolumn{1}{l}{}         & \multicolumn{1}{l}{}         \\
                      & \multicolumn{4}{c}{$M_\mathrm{bin}\in[10^7,10^8] M_\odot$}                                                                \\ \cmidrule(l){2-5} 
All mergers           & \multicolumn{2}{c}{$(4.68\pm3.09)\times10^{-4}$}            & \multicolumn{2}{c}{~~~~~~~~0.18$\pm$0.02}                           \\
SNR\textgreater{}8    & $(1.81\pm2.17)\times10^{-4}$ & $(4.68\pm3.09)\times10^{-4}$ & $(6.48\pm1.77)\times10^{-3}$ & 0.18$\pm$0.02                \\
SNR\textgreater{}100  & $(1.77\pm0.78)\times10^{-5}$ & $(4.68\pm3.09)\times10^{-4}$ & $(1.09\pm0.25)\times10^{-4}$ & 0.15$\pm$0.02                \\ \bottomrule

\label{table:merger_rate}
\end{tabular}
\end{table*}

\section{Summary and Conclusions}
\label{sec:conclusions}

We have investigated the formation and merger rate of IMBH and SMBH binaries, focusing on their detectability by the upcoming gravitational-wave detectors LGWA and LISA. We generated the binary population using a light seed scenario consistent with Pop III seeding, applying the CAM25 model to merger trees from a \textsc{pinocchio} simulation with a 40 Mpc/h box. The CAM25 model was then used to estimate the baryonic properties of the host galaxies and the growth of the BHs through mergers and accretion. From the halo merger redshifts provided by the simulation, we calculated the BH merger redshifts by adding two time delays: from halo to galaxy merger, and from galaxy merger to BH merger. To calculate the first delay, we use the prescription in CAM25. For the second delay, we considered only the timescale of DF. 

To estimate the second delay, we applied the Chandrasekhar prescription to compute the time taken by the secondary BH to sink towards the centre. We considered two limiting cases of this delay: one where the secondary BH is completely stripped of all surrounding stellar material for the entire plunge, and another where the entire stellar material of the secondary galaxy remains intact for the entire timescale. The former gives an upper limit on the timescale, while the latter gives a lower limit. 

After calculating the DF timescales for each pair of BHs, we constructed the populations of merging BHs. We then randomly distributed all the events in the sky and evaluated their SNRs as seen by the LGWA and LISA detectors. Below, we summarise our main findings about the detected population, along with the estimated timescales:

\begin{itemize}
    \item In the lower limit scenario for the DF timescale, the number of mergers completing within a Hubble time is significantly larger than in the upper limit case. Out of the entire population of the BH pairs, only 2\% have delays less than the Hubble time in the upper limits case, whereas nearly 40\% have shorter delay in the lower limits case. The difference mainly comes from using the entire stellar mass of the secondary galaxy in evaluating the delay instead of just the BH mass. This reduces the delay by a few orders of magnitude and highlights the importance of accurately modelling galaxy stripping to estimate BH merger timescales.

    \item The larger number of mergers completing within the age of the universe in the lower limit case defines our optimistic merger rate scenario, while the upper limit case represents a pessimistic scenario. In both cases, the majority of mergers are major, with mass ratios typically exceeding 0.01.

    \item The redshift distribution of mergers strongly depends on the assumed timescale: in the optimistic scenario, BH mergers can occur as early as $z\sim14$, while in the pessimistic case they mostly occur at $z<8$. The difference in rates among the two cases is higher at higher redshifts, while it reduces at lower redshifts down to only around an order of magnitude at $z=0$. Despite these differences, most detectable events occur at lower redshifts ($z<3$).

    \item We evaluated the GW signal strength of each merger and assessed its detectability with LISA and LGWA. Our results show that both LGWA and LISA can detect around 25 mergers per year in the optimistic case including all the BH masses. However, that number reduces to around 0.5 per year in the pessimistic case. 

    \item For the 10 year expected lifetime of LGWA, our model predicts that it will be able to detect a total of $\sim260$ mergers in the optimistic case, while only $\sim5$ in the pessimistic case. On the other hand, LISA is expected to be active for 4 years, yielding a total of $\sim105$ and $\sim2$ merger detections in the optimistic and pessimistic cases respectively.

    \item Considering the binaries with total mass in the intermediate range, the number of detections per year for LGWA ranges from $\sim0.34$ in the pessimistic case to $\sim20$ in the optimistic case. LISA detects slightly less, ranging from $\sim0.3$ to $\sim15$ for both the cases respectively. For the number of detections with higher SNR, LGWA has a higher detection rate, as expected. However, detections with SNR greater than 100 are rare, with only around $\sim1$ per year for LISA, and around $\sim2.7$ for LGWA, both in the optimistic case.

    \item For the binaries in the range of $10^5 M_\odot$ to $10^7 M_\odot$, the detection tendency flips, and LISA is able to detect more than $\sim11$ per year in the optimistic case, compared to $\sim7$ for LGWA. For binaries detected with SNR > 100, LISA is still able to detect more than $\sim6$ mergers per year in the optimistic case, compared to less than one for LGWA. For the most massive binaries with $M_\mathrm{bin}$ higher than $10^7~M_\odot$, the rates are quite low which is expected since these BHs lie outside the design sensitivity of the detectors.

    \item The merger rates obtained from our model for LISA are consistent with the rates mentioned above and with the literature, despite the differences in the modelling and the simulations.
\end{itemize}

Finally, we note that in our calculation of the DF timescale, we only considered the variation in the mass of the infalling object which resulted in a difference of orders of magnitude. Due to this huge variance, we expect that other assumptions, particularly the assumption of initial separation as the radius of the remnant (or primary) galaxy, to not make a significant difference in the calculation of this timescale. We also note that we only followed the formation of binaries through the galaxy merger channel in this work. This means that we do not account for binaries formed through other channels, including the formation of IMBHs through runaway stellar mergers or the formation of Intermediate or Extreme Mass Ratio Inspirals. These binaries can also be detected by LGWA and LISA, which implies that the total detection rate of BH binaries could be even higher than the results presented here. Nevertheless, both detectors offer complementary sensitivity across different mass and redshift regimes. The combination of LISA and LGWA will be essential to probe the full BH mass spectrum, from the elusive intermediate-mass to the supermassive regimes, from major mergers with high mass ratio to extreme mass ratio inspirals. Furthermore, due to the dependence of the merging redshifts and the number density of mergers on the formation and seeding mechanisms of BHs, the detections of their mergers will help in constraining those mechanisms as well.

\begin{acknowledgements}
We thank the anonymous referee for the useful and constructive comments which improved the quality of the paper. We acknowledge a financial contribution from the Bando Ricerca Fondamentale INAF 2022 Large Grant, ‘Dual and binary supermassive black holes in the multi-messenger era: from galaxy mergers to gravitational waves’ and from the INAF Bando Ricerca Fondamentale INAF 2024 Large Grant, `The Quest for dual and binary massive black holes in the gravitational wave era'. We also acknowledge the support of the computing centre of INAF-Osservatorio Astronomico di Trieste, under the coordination of the CHIPP project \cite{Bertocco20, Taffoni20}. JS thanks David Izquierdo-Villalba, Elisa Bortolas, and Manuel Arca Sedda for helpful discussions on DF and the referee report. VC thanks the BlackHoleWeather ERC project and PI Prof. Gaspari for salary support. JCT acknowledges support from ERC Advanced Grant 788829 (MSTAR). 
\newline
The data underlying this article will be shared on reasonable request to the corresponding author.
\end{acknowledgements}

\bibliographystyle{aa} 
\bibliography{bibliography}

\end{document}